%% file: main.tex
\documentclass[a4paper,11pt]{article}
\usepackage{pos}
\usepackage{makecell}

\newboolean{uprightparticles}
\setboolean{uprightparticles}{false}

\input{lhcb-symbols-def}

\title{Selecting long-lived particles in the first trigger level at the LHC}

\author*{Lorenzo Pica}

\onbehalf{for the LHCb collaboration}

\affiliation{INFN Sezione di Pisa,\\
  Largo Bruno Pontecorvo 3, Pisa, Italy}

\affiliation{Scuola Normale Superiore,\\
Piazza dei Cavalieri 7, Pisa, Italy}

\emailAdd{lorenzo.pica@cern.ch}

\abstract{The LHCb experiment is starting to take data in Run 3 with a new DAQ  system, capable of performing complete event reconstruction at the full LHC collision rate. One  novel  opportunity  offered  by  this  system  is  triggering on long-lived particles (LLPs) at the very first stage of the trigger. This could potentially increase trigger efficiency for LLPs, typically suffering from low online detection efficiency at hadron collider experiments, because of their decay signatures. We investigated the feasibility and effectiveness of an early LLP-triggering approach in LHCb with the implementation of two LLP-dedicated selections in the first trigger level (HLT1), targeting the presence of either one, or two, \KS decays. Selection tuning is performed on simulation, targeting some benchmark channels with \KS particles in the final state, as \Dz\to\KS\KS and \Bz\to\KS\KS. Tests ran on simulated samples predict a large increase in selection efficiency, up to 2.6x for the \Dz\to\KS\KS channel, at the price of a very modest increase of HLT1 computational load and trigger rate. These selections were implemented in the GPU-based HLT1 trigger sequence, and took data during the physics data-taking LHCb run in year 2022. In this document, we present results obtained from these first data, yielding good quality \KS and \KS-pair samples even from a very limited integrated luminosity. We conclude with a discussion of the physics prospects opened by these new triggers, and their planned extension to tracks decaying outside the volume of the VELO subdector (”downstream tracks”) to further extend their acceptance.}

\FullConference{%
  Corfu2023 Workshop on Future Accelerators\\
  23-29 April 2023\\
  Faliraki, Old town Corfu, Greece
}

\tableofcontents

\begin{document}
\maketitle
\raggedbottom
\section{LLP scenario}
Many measurements in High Energy Physics (HEP) involve particles decaying with extremely short flight distances, as the electroweak bosons $W^{\pm}/Z^0$ and the Higgs boson $H^0$. \textit{Beauty} and \textit{charm} hadrons have longer lifetimes ($\tau \sim 10^{-12}\sec$), but their flight distance is still limited to few \mm, even in multi-TeV collisions.
This common feature drove the design of most general-purpose experiments, aimed at detection and selection of short-lived states.
However, particles with longer flight distances, up to hundreds of centimeters, usually referred to as long-lived particles (LLPs) are also involved in a large number of interesting processes.

In many cases, the sensitivity of measurements involving LLPs suffers of specific experimental limitations  related to their long flight paths. In particular, at hadron colliders one of the important issues is the difficulty of achieving a good efficiency in the online selection of such events.
However, the advent of full-event reconstruction in real time opens new possibilities also in this respect. In this work, I will show that it can be exploited to significantly increase the LLP online selection efficiency at LHCb.
Before going through this, it is worth mentioning some notable LLP measurements, motivating these efforts.

\subsection{Beyond Standard Model LLPs physics case}
Among the particles predicted by BSM theories, several have a large lifetime, because of the very small coupling between BSM state and detectable SM particles, or a reduced available phase space for the BSM particle decay. The search for such LLPs is therefore of extreme interest, since any detection would be an evidence of BSM physics.

Supersimmetry (SUSY) is still one of the most interesting BSM frameworks. It predicts the existence of a supersymmetric partner for every SM particle and some of these are expected to have a large lifetime, behaving experimentally as LLPs. Searches for these have been extensively performed and taking the ATLAS collaboration as an example, several decay signatures have been analyzed: fully-hadronic displaced vertex with an associated muon~\cite{ATLAS:2022fag}, a single high-impact-parameter lepton~\cite{ATLAS:2020wjh}, displaced vertices when collisions are absent~\cite{ATLAS:2021mdj}, so-called \textit{disappearing tracks}~\cite{ATLAS:2022rme}, or multi-charged particles~\cite{ATLAS:2023zxo}.

A possible production mode for BSM particles is through the decay of an Higgs boson (the so-called Higgs portal).
This process is particularly interesting as it is allowed by the current knowledge on the Higgs boson total decay width, possibly including decays into unknown objects with branching fractions up to $30\%$~\cite{ATLAS:2016neq}. Searches have been performed by different collaborations, as D0~\cite{D0:2009mtx}, CDF~\cite{CDF:2011dnt}, ATLAS~\cite{ATLAS:2015mlf}, CMS~\cite{CMS:2014wda} and LHCb~\cite{LHCb:2017xxn}.
Taking as example the latest LHCb result, a search is performed for the hidden sector lightest state $\pi_{\nu}$, exploiting 2\invfb of data collected during Run 1 of LHC. Considered mass and lifetime hypothesis lay in the $25-50\gevcc$ and $2-500$ ps ranges, respectively. The presence of a displaced vertex having transverse distance from beam axis $R_{xy}$ larger than $0.4\mm$ and two associated hadronic jets is required during online selection.
No evidence has been found for $\pi_{\nu}$ during analysis and upper limits on the cross section are set.

Heavy neutral leptons (HNLs) are another example of BSM LLPs. These are heavy counterparts of SM neutrinos by the so-called Seesaw mechanism~\cite{Mohapatra:1979ia, Yanagida:1979as, Schechter:1980gr}, with their mass upper end being $O(\tev/c^2)$.
Several searches for HNLs have been performed and an extensive review can be found in Ref.~\cite{Abdullahi:2022jlv}. An interesting example is the HNLs search recently published by CMS collaboration, based on the full statistics collected during Run 2 of the LHC (138\invfb)~\cite{CMS:2022fut}.
Here the presence of three leptons in the final state is required - one coming from the primary proton-proton interaction, and the others forming a displaced secondary vertex within the CMS tracking system. No evidence for BSM decays is found and upper limits are set on the mixing probability between SM and BSM neutrinos.

Many other BSM theories predict the existence of LLPs, detectable in current and future experiments, as dark photons~\cite{Graham:2021ggy} or axion-like particles~\cite{Hook:2019qoh} -- it is beyond the scope of this document to discuss all of them. The examples mentioned should be sufficient to convey the richness of BSM physics accessible through LLPs.
\subsection{Standard Model LLPs physics case}
SM LLPs play a crucial role in current SM experimental physics as well. They give access to a large number of interesting measurement and,  additionally, they offer a perfect ground for  developing and testing LLPs-specific experimental techniques.  They are abundantly produced and detected already in present experiments, with strange particles \KS and \Lz making the most common examples.

The measurement of the \KS\to\mup\mun decay branching ratio is an outstanding example of the impact of measurements involving SM LLPs. A precise knowledge of this parameter is interesting because of the very small value of the SM prediction, being $5\times 10^{-12}$~\cite{DAmbrosio:2017klp}, and making it an excellent null test for the SM, sensitive to new physics influence. BR(\KS\to\mup\mun) has been measured several times, and \lhcb published the best and most recent result~\cite{LHCb:2020ycd}, exploiting data collected during Run 2 of the LHC ($5.6\invfb$).
Despite the huge amount of produced \KS mesons, no evidence for this decay has been found, and an upper limit is set to $BR(\KS\to\mup\mun) < 2.1 \times 10^{-10}$ at $90\%$ CL~\cite{LHCb:2020ycd}.

CP violation (CPV) in \textit{charm} decays is another phenomenon that can be investigated through the study of \KS. It has been observed for the first time by LHCb collaboration in 2019~\cite{LHCb:2019hro}, analyzing \Dz\to\Kp\Km and \Dz\to\pip\pim decays. However, agreement with the theory can't currently be determined, because of predictions precision level. The observation of additional charm CPV manifestation can improve experimental precision and provide a crucial input to theory. \Dz\to\KS\KS represent an excellent candidate for a charm CPV confirmation, since its time-integrated CPV size could be enhanced up to $1.1\%$~\cite{Nierste:2015zra}. Several measurements of \mbox{\ACP(\Dz\to\KS\KS)} has been performed \cite{CLEO:2000opx, Dash:2017heu, LHCb:2015ope, LHCb:2021rdn}, and thanks to the most recent LHCb result~\cite{LHCb:2021rdn} world average approached theoretical predictions upper end for the first time, strongly motivating the collection of additional statistics.

Also the system of beauty neutral mesons offer an important measurement when LLPs are involved, with the \Bz\to\jpsi\KS decay, providing a theoretically clean access to the CKM angle $\beta$~\cite{Carter:1980tk, Bigi:1981qs} through the measurement of its CPV parameters. Several measurements have been performed ~\cite{BaBar:2009byl, Belle-II:2023nmj, LHCb:2023zcp} and with the current precision no tension has been found in the determination of the CKM angles, motivating a more precise $\beta$ determination.
\subsection{LLPs study at the LHC}
LLPs measurements are performed at different experiments and facilities. Results from \atlas, \cms, \lhcb, \babar, and \belle have been reported, but also other are involved in LLPs study, as non-accelerators ones (IceCube~\cite{IceCube:2002eys} and Super-Kamiokande~\cite{Super-Kamiokande:2002weg}) or LLP-dedicated experiments, both present (NA62~\cite{NA62:2017rwk}) and future ones (CODEX-b~\cite{Aielli:2019ivi} and FASER~\cite{FASER:2019aik}).
In this wide scenario the Large Hadron Collider is one of the most suitable facilities for the study of LLPs.
In fact, thanks to the very large cross section provided by hadron collisions, it allows the collection of huge samples in a very short timescale, with the most advanced HEP detectors recording produced collisions. Additionally, LHC, and CERN in general, can profit from an almost-unique decade-long timescale for their facilities and projects, as stated in the last update of the European Strategy for Particle Physics~\cite{fcc_ee}.

However, the large LLP production rate made available by the LHC is just the first step toward the collection of the large statistics required to achieve high-precision measurements. In fact, in order to be exploited for offline analyses, LLPs need to be identified in real time and saved to permanent storage. This task is fulfilled by the so-called \textit{trigger} algorithm, performing real-time selection of interesting processes.
\subsection{LLP triggering}
Triggering is an extremely challenging task at the LHC, because of the very crowded environment of hadron collisions and the very small time window in which the algorithm has to operate. Because of these tight constraints, signal identification is usually based on very distinctive signatures, as a large transverse momentum/energy deposit. These provide an effective handle for signal identification and are also accessible with the exploitation of fastest sub-detectors, as calorimetric and muon systems.

However, this strategy has a limited performance when applied to LLPs decays, being their typical decay signatures displaced tracks and vertices, characterized by a relatively low \pt, significantly different from the ones of prompt objects. The implementation of a dedicated trigger strategy is extremely challenging, since the identification of such signatures requires a series of computationally heavy tasks, as tracking systems read-out and tracks and vertices reconstruction. Because of this, LLP-dedicated triggers are typically implemented only in later stages of the trigger, when more processing time is available.
However, this condition strongly limits the efficiency achievable in the early stages of the trigger, where LLP decays were retained only when an unrelated process randomly fired the trigger. It can therefore be expected that if the LLPs selection could be moved to  earlier stages of the trigger, their online selection efficiency could improve significantly. LHCb, with the implementation of its upgraded detector and DAQ system, is a very  suitable LHC experiment for attempting this.

\section{LLP triggering at LHCb}
LHCb is a forward single-arm spectrometer, specifically designed to study \textit{beauty} and \textit{charm} hadron decays. It profits from an excellent momentum resolution and vertex identification precision with its tracking system~\cite{LHCb:2023hlw}, \pion/\kaon/\proton separation with its Particle IDentification system (PID)~\cite{LHCb:2023hlw} and muon identification through a dedicated sub-detector~\cite{LHCb:2023hlw}. Its acceptance covers the $2<\eta<5$ range\footnote{Pseudorapidity \Peta, defined as $-\log(\tan(\theta/2))$, where $\theta$ is the angle between the beams flight direction and the momentum of the particle.}, making it complementary to other LHC major experiments.
LHCb went under a major upgrade during the long shutdown 2 of the LHC (so-called Upgrade I~\cite{LHCb:2023hlw}) that has seen the complete replacement of its entire tracking, DAQ and trigger systems. The aim is to achieve real time reconstruction of tracks and vertices at the full LHC bunch-crossing rate.

\subsection{Track reconstruction in LHCb}
The upgraded LHCb tracking system is composed of three sub-detectors. The VErtex LOcator (VELO) is a silicon pixel detector, providing precise track and vertex reconstruction. The Upstream Tracker (UT) is an additional silicon detector placed upstream the magnet, speeding up event reconstruction and reducing fake tracks reconstruction rate. The Scintillating Fibre (SciFi) is placed downstream the magnet and allows the measurement of particles momentum.

Particle trajectories (\textit{tracks}) can be reconstructed inside the LHCb detector with the entire tracking system, or just a part of it, and are classified as:
\begin{itemize}
    \item \textbf{Long tracks}, reconstructed exploiting VELO, UT and SciFi - these offer the best resolution;
    \item \textbf{Downstream tracks}, reconstructed exploiting UT and SciFi only - these offer a worse resolution, but a larger acceptance for LLP decays;
    \item \textbf{T-tracks}, reconstructed exploiting SciFi hits only.
\end{itemize}
\subsection{The LHCb trigger architecture}

LHCb employs a fully software-based trigger system in Run 3, made of two subsequent steps, named HLT1 and HLT2. It implements a trigger-less readout of the entire tracking, calorimetric and muon detectors.
HLT1 is designed to perform an event rate reduction of a factor of 30, with an output event rate of 1\mhz. This is achieved running a 30 MHz read-out and reconstruction of tracks and vertices.

This innovative approach enables the implementation of LLPs trigger selections at the earliest stage of the trigger, a first time ever at an hadron collider.
This possibility has been investigated, designing HLT1 LLP-dedicated selections, in order to understand their sustainability in term of throughput and background rejection and the potential achievable efficiency gain. Only objects decaying within \velo acceptance will be considered at this stage, since only long tracks are reconstructed in the HLT1 design configuration.

\subsection{New trigger lines}
Given the novelty and complexity of this task, \KS decays have been chosen as targets of novel selections, in order to work with a particle whose behavior is well-known and easily reproducible on simulation. Additionally, since \KS mesons are hugely produced in LHC collisions, any commissioning operation of novel selections would be significantly simplified and sped up.

Two different selections (trigger \textit{lines}) have been designed and implemented. \twotrks targets single \KS candidates and \twoks selects \KS candidates pairs, without any requirement on the parent particle.
Lines are based on a set of rectangular selections. The exploitation of any classifier is excluded, in order to avoid the introduction of any correlation between exploited variables, potentially limiting precision on performed measurements. Additionally, this approach provides an easier understanding of selection impact, crucial in a commissioning phase.
\subsubsection{Novel lines design and simulation studies}
The computational cost of any new HLT1 algorithm must be carefully guarded, in order to maintain the required 30 MHz processing frequency. The novel lines we introduced satisfy this, having almost no impact on HLT1 throughput by design. In fact, displaced vertices reconstruction is run independently from \KS-dedicated lines, and it anyway represents just a small fraction of overall timing ($O(\%)$), and \twoks line does not require any additional vertex reconstruction, since no selection is applied on \KS parent.

Any novel HLT1 selection must also achieve high acceptance on target decays while maintaining an excellent background rejection, since output bandwidth is almost fully allocated for the selection of heavy flavor decays~\cite{LHCb:2021kxm}.
In order to meet such tight requirements, thresholds has been tuned exploiting the TMVA rectangular cuts optimizer~\cite{Hocker:2007ht}. \KS candidates present in a \mbox{\Dz\to\KS\KS} simulated sample have been adopted as signal sample. Any two-track combination with an invariant mass within $\pm 70\mevcc$ from $m(\KS)$ present in a simulated sample of generic \textit{pp} collisions (\textit{minimum bias}) has been adopted as background sample. Variables exploited in the selections are:
\begin{itemize}
    \item \chisqndf: \chisq obtained from the fit of a track reconstructed in the detector, normalized to its degrees of freedom;
    \item \Pp/\pt: momentum/transverse momentum of the particle;
    \item \textit{Impact Parameter} (\textit{IP}): distance of closest approach between the PV and the direction identified by the momentum of the particle;
    \item \chisqip: difference between \chisq obtained from PV fitting including a particle in the fit or not;
    \item \chisqvtxndf: \chisq of tracks origin vertex fit, normalized to its degrees of freedom;
    \item $\theta_{DIRA}$: angle between the momentum of a particle and the direction identified by the PV and the decay vertex of that particle;
    \item $\theta_{\pion\pion}$ ($direction\;angle$): angle between the two particles forming the a decay vertex;
    \item IP combination: $IP(\pip)\times IP(\pim) / IP(K^0_S)$ ratio between daughters $IP$ product and parent $IP$.
\end{itemize}
However, only selections on $\chisqip(\pi)$, $\pt(\pion)$, $p(\pion)$, $\pt(\KS)$ and $IP$ combination have been numerically optimized to achieve maximum signal/background separation, while remaining are kept fixed during optimization. These have been set following a different rationale, as ensuring a correct background shape identification in $m(\pip\pim)$ sidebands, or avoiding the application of a too tight selection in case of data/simulation disagreement for $\chisq_{vtx}(\KS)$, $cos(\theta_{DIRA})$ or $cos(\theta_{\pi\pi})$ (especially during commissioning phases).

Selections defining \twotrks and \twoks lines are reported in Table \ref{tab:twotrks_twoks_sel}.
\begin{table}[htp!]
\begin{center}
\begin{tabular}{c|c|c}
\hline
Variable    & \twotrks selection & \twoks selection \\
\hline
\hline
$\chisq/ndf$         & $<2.5$ & $<2.5$\\
\hline
$\chisqip(\pi)$          & $>50$ & $>15$\\
\hline
$\pt(\pion)$ &  $>470 \mevc$ & $> 425\mevc$\\
\hline
$p (\pion)$ & $>5 \gevc$ & $> 3\gevc$\\
\hline
$|m(\pip\pim) - m(\KS)|$ & $<45\,\mevcc$ & $<45\,\mevcc$ \\
\hline
$\chisq_{vtx}(\KS)$      & $<20$ & $<20$ \\
\hline
$\eta(\KS)$ & $2 < \eta(\KS) < 4.2$ & $2 < \eta(\KS) < 4.2$ \\
\hline
$\pt(\KS)$ & $> 2500 \mevc$ & $>1150 \mevc$ \\
\hline
$cos(\theta_{DIRA})$ & $ > 0.99$ & $ > 0.99$ \\
\hline
$cos(\theta_{\pi\pi})$ & $ > 0.99$ & $ > 0.99$ \\
\hline
$\frac{IP(\pip)\times IP(\pim)}{IP(K^0_S)}$ & $> 0.72$ mm & $> 0.23$ mm \\
\hline
\end{tabular}
    \caption{Selections applied by \twotrks and \twoks HLT1 line. In case of \twoks line two \KS candidates satisfying reported selections are required in the same event.}
    \label{tab:twotrks_twoks_sel}
\end{center}
\end{table}
Specific \KS signatures are exploited to keep background acceptance under control, as the specific invariant mass and a significant \pion displacement, while other are loosened to increase efficiency, as \pt(\pion). \twoks line applies lower thresholds for single \KS candidate since requirement for a candidates pair reduces background acceptance.

The impact of the new lines on the HLT1 background acceptance has been estimated with the variation of HLT1 \textit{rate}, hence HLT1 output rate when processing a 30 MHz minimum bias input. \twotrks and \twoks respectively determine a $40\khz$ and $11\khz$ increase of the overall HLT1 rate, corresponding to a variation limited to $4\%$ and $1.1\%$.

The efficiency has been evaluated over a set of representative decays. It is computed as the ratio between the number of decays retained by a line and the total number of decays whose final state tracks can be reconstructed through long tracks.
The \twotrks line determines the most significant gain in case of \mbox{\Dz\to\KS\KS} decays, where HLT1 efficiency increases by a factor of $2.6$. Other decay channels benefit from its introduction as well, as a $20\%$ increase is estimated in case of \mbox{\Bz\to\KS\KS} (reaching $77\%$) and a $45\%$ gain is estimated for \mbox{\Bz\to\KS\piz}.
\twoks line shows an efficiency on \Dz/\Bz\to\KS\KS decays larger than general-purpose track-based lines (that occupy more than $80\%$ of the HLT1 output bandwidth). However, it shows a lower efficiency w.r.t. \twotrks, despite looser thresholds. This is expected, since the minimum \pt(\pion) required for a track in order to be reconstructed in HLT1, determines a tighter selection when four tracks are requested in the final state.
\section{Initial commissioning}
Performances estimated on simulation motivated \twotrks and \twoks lines implementation in the HLT1 system, allowing their exploitation at the start of Run 3 data taking, in 2022. Data collected during this phase allowed a crucial early commissioning of novel lines, even if collected during a commissioning phase of detector and trigger system. This is essential in order to take prompt action against any deviation from expected performances.

Lines performance on real data has been investigated analyzing a small sub-sample of 2022 data, corresponding to an integrated luminosity of $240\invnb$.

\subsection{Candidate invariant mass distributions}
\begin{figure}[h]
    \centering
    \includegraphics[scale=0.3]{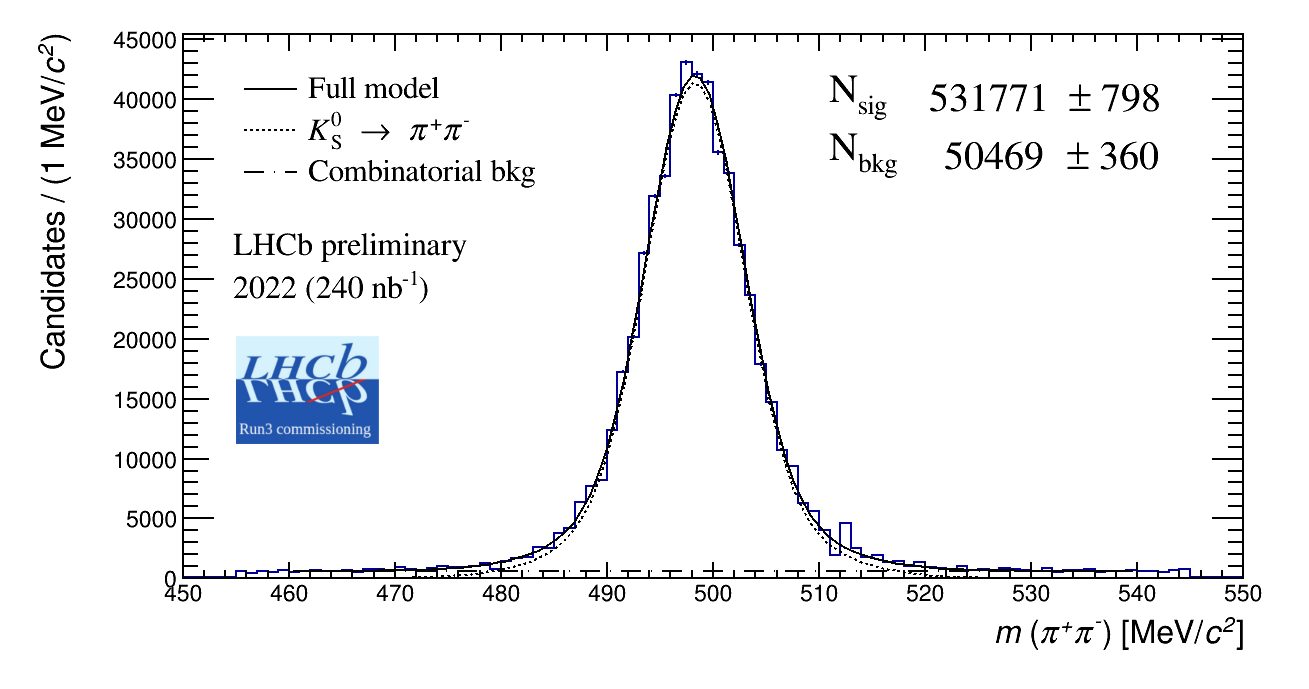}
	\caption{Invariant mass for \KS candidates and collected by \twotrks. Taken from~\cite{LHCB-FIGURE-2023-005}.} 
	\label{fig:m_twotrks_commissioning}
\end{figure}

The first  performed check is the inspection of \KS candidates invariant mass distribution, reported in Figure \ref{fig:m_twotrks_commissioning} for the \twotrks line.
A peak is clearly present, centered around the known $m(\KS)$, with a gaussian shape and a very modest background contribution.

\begin{figure}[h]
	\centering
    \includegraphics[scale=0.3]{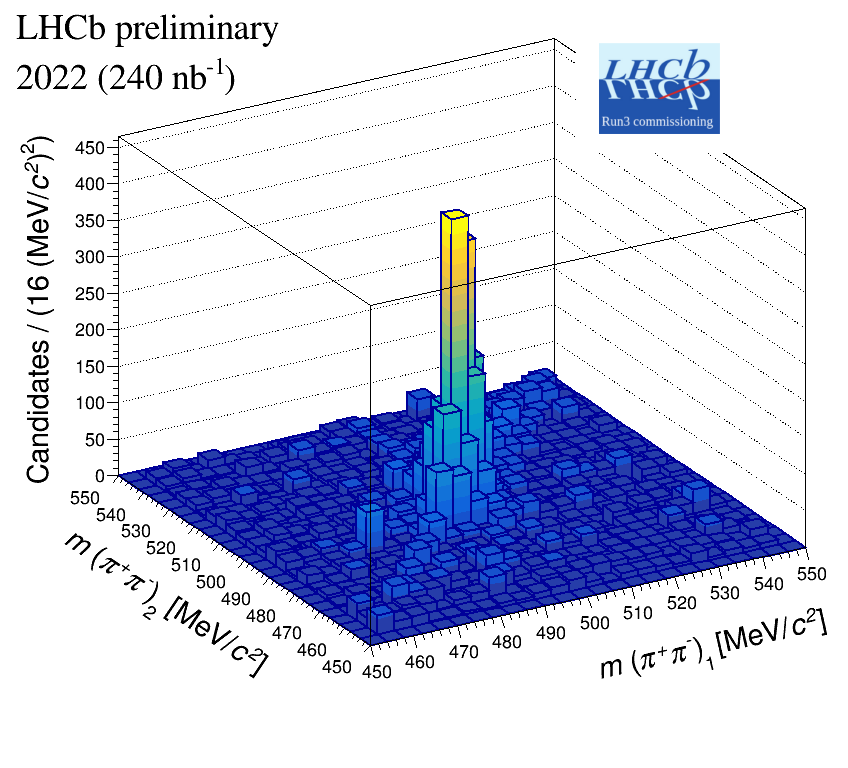}
	\caption{Invariant mass for \KS candidate pairs collected by \twoks. Taken from~\cite{LHCB-FIGURE-2023-005}.} 
	\label{fig:m_twoks_commissioning}
\end{figure}

The 2D \KS candidates invariant mass distribution for candidates collected by \twoks line is reported in Figure \ref{fig:m_twoks_commissioning}.
Clear evidence for a peak due to real \KS pairs is present in this case as well, centered around $m(\KS)$ for both candidates. \twoks line collected $\sim 2.5$k candidates, $\sim 25\%$ of which are contained in the peak.

Considered distributions show how novel HLT1 lines successfully selected a sizeable sample of \KS candidates and \KS candidate pairs in \textit{pp} collisions at the full LHC crossing rate of 30 MHz, a result never achieved before at an hadron collider.

It has been decided not to try to evaluate the efficiency of the lines on the available commissioning data, since a reliable determination would be hard to achieve, because of the preliminary detector performances, significantly overlapping with selection ones.
\subsection{Rates on real \textit{pp} collisions}
Reliably reproducing generic \textit{pp} collision on simulation is a challenging task and new selections background acceptance could be significantly higher on real data than expectations, potentially making lines exploitation unfeasible. Therefore, a crucial check is the estimate of lines rate on real data.

Two unfiltered samples have been exploited, collected at different instantaneous luminosity values, here quantified through the parameter $\mu$, the average number of visible \textit{pp} interactions per bunch crossing.
Rates computed at different $\mu$ values are reported in Table \ref{tab:rates}.
\begin{table}[htp]
    \centering
    \begin{tabular}{ccc}
        $\mu$ & Rate \twotrks [kHz] & Rate \twoks [kHz] \\
        \hline\hline
        1.1 & $1.24 \pm 0.08$ & $0.05 \pm 0.02$ \\
        \hline
        3.1 & $4.2 \pm 0.5$ & $2.5 \pm 0.4$ \\
        \hline
        \thead{5.32\\(Simulation)} & $74 \pm 7$ & $13 \pm 3$ \\
        \hline
    \end{tabular}
    \caption{\twotrks and \twoks HLT1 lines rates computed on real data and simulation.}
    \label{tab:rates}
\end{table}
Measured rates appear to be well under control, largely below simulations predictions for both considered $\mu$ values, despite the growth present when considering higher $\mu$ value. The \twoks line rate shows a steeper rate increase, as expected for a 4-track selection. Data at the nominal LHCb Run 3 luminosity ($\mu=5.5$) are not available, preventing from a direct comparison with simulation predictions. Nevertheless, this test clearly verifies finiteness of implemented selections rate, demonstrating their feasibility and motivating their exploitation also during Run 3 physics data taking.
\section{Possible acceptance extension}
The early triggering LLP-dedicated selections here considered demonstrated a significant impact in LLP online selection efficiency. However, their effect is anyway constrained by a major acceptance limitation, hence lack of downstream and T tracks reconstruction at HLT1 level, preventing retention of the significant fraction of LLP decaying outside of the VELO acceptance.
\subsection{HLT1 downstream triggering potential}
The potential efficiency improvement achievable in case of downstream triggering availability is estimated. A simplified offline HLT1 emulator made of three lines is exploited. Considered lines are: a single-track (1-track) and a two-track (2-track) selection emulating general-purpose LHCb HLT1 lines and a \KS-dedicated one, implementing \twotrks selections. The latter is the only one extended to downstream triggering.
In order to estimate efficiency gain and rate cost of acceptance extension, the considered system is run over a set of samples (both real and simulated data). These are reconstructed with an offline-level precision, higher than the one achievable during HLT1 processing. However,
it has been decided to ignore this difference, since it is expected to do not have any significant impact, being this a first order estimate.

Efficiency gain when \twotrks line is extended to downstream triggering has been estimated on some representative \KS-final state decays, results are reported in Table \ref{tab:eff_down}.
\begin{table}[htp]
    \centering
    \begin{tabular}{cc}
        \hline
         Sample & Efficiency gain factor \\
         \hline
         \hline
         \Dz\to\KS\KS & 3.3 \\
         \hline
         \Dz\to\KS\pip\pim & 1.14 \\
         \hline
         \Bz\to\KS\KS & 5.6 \\
         \hline
         \Bz\to\KS\pip\pim & 1.09 \\
         \hline
    \end{tabular}
    \caption{Efficiency gain achieved extending \twotrks HLT1 line to downstream reconstructed LLP candidates.}
    \label{tab:eff_down}
\end{table}
Fully \KS-final state decays show the most significant improvement, reaching a factor larger than $5$ ($3$) in case of \mbox{\Bz\to\KS\KS} (\mbox{\Dz\to\KS\KS}). It's worth noting how these gains add up to the ones due to the implementation of \KS-dedicated HLT1 line triggering on long tracks only. Therefore, the implementation of a \twotrks-like line, without any acceptance limitation, could determine an efficiency improvement up to a factor $8.5$, when considering \mbox{\Dz\to\KS\KS} decays.
A less significant efficiency improvement is achieved in case of \KS\pip\pim final state decays. This is expected, since the \pip\pim pair directly coming from the \Dz efficiently triggers general-purpose single- and two-track selections and the fraction of decays triggered through the \KS is lower in any case. However, in this case triggering on the final state \KS could prevent the insurgence of major systematic effects during offline analysis, appearing when events are triggered through the \pip\pim pair. In this scenario, having the possibility of triggering also on \KS decaying outside \velo is a key factor in order to collect a sizeable statistics sample and the verified efficiency gain is anyway important.

The rate cost of considered downstream trigger configuration is determined exploiting the same HLT1 emulator, ran over a sample of \textit{pp} collisions collected by LHCb in 2018, scaling estimated rates to Run 3 luminosity. Real data exploitation is preferred in this case, in order to have a more reliable estimate.
Results are reported in Table \ref{tab:rate_down}.
\begin{table}[htp]
    \centering
    \begin{tabular}{cc}
        \hline
        Selections & Rate [kHz] \\
        \hline
        \hline
        1-track $\vee$ 2-track & $1518 \pm 92$ \\
        \hline
        \thead{1-track $\vee$ 2-track\\ $\vee$ \twotrks \\(long tracking)} & $1518 \pm 92$ \\
        \hline
        \thead{1-track $\vee$ 2-track\\ $\vee$ \twotrks \\(long and downstream tracking)} & $1767 \pm 99$ \\
        \hline
    \end{tabular}
    \caption{Rate estimates in case of downstream tracking availability at HLT1 level.}
    \label{tab:rate_down}
\end{table}
The rate of system given by 1-track and 2-track is of the expected order of magnitude ($O \sim 1\mhz$), even if at first order. The addition of \twotrks does not cause any significant increase in the rate at the considered precision level, agreeing with the minimal impact verified on simulation and real data.
Rates for these two configurations have been computed as a sanity check. Achieved results suggest a reliable emulation of HLT1 system, indicating the same also when considering downstream tracks.
Extension of \twotrks line to downstream triggering determines a limited rate increase, corresponding to $15\%$ of the overall HLT1 system rate. This result strongly suggests feasibility of considered approach, representing a crucial step toward downstream tracking implementation.
\subsection{Downstream tracking efforts}
Efforts toward the implementation of HLT1 downstream tracking are already ongoing within the LHCb collaboration. Standalone T-tracks reconstruction has been already ported to HLT1, implemented in GPU framework~\cite{Calefice:2022upw}, and exploited during data taking.
However, this task is very computationally expensive, occupying $\sim45\%$ of the overall HLT1 timing budget~\cite{Calefice:2022upw}, strongly motivating efforts aimed at reducing this time.

One of the approaches considered in order to speed up T-tracks reconstruction in future LHC runs involves the implementation a dedicated device, performing reconstruction of T-tracks at readout level, before HLT1, investigated by LHCb with an advanced R\&D project, named DoWnstream Tracker (DWT)~\cite{Cenci:2020fbz}. It targets a 30 MHz forward tracking (SciFi/UT-SciFi) from Run 4, exploiting the so-called RETINA algorithm, implemented on a FPGA-based device. Forward tracks reconstruction is performed before event-building, making track appear as raw data coming out of detector. The implementation of such a device allows HLT1 downstream triggering, and the achievement of aforementioned efficiency gain, while saving a significant amount of processing time both at HLT1 and HLT2 level, making room for the execution of other tasks, as event reconstruction at an earlier trigger stage or the implementation of more channel-specific selections.

\section{Conclusions}
We presented results from a first example of LLP-dedicated selections working at the earliest trigger level at LHC, targeting the selection of \KS candidates.
Promising simulation studies motivated the implementation of these selections in LHCb trigger system and their exploitation at the re-start of LHCb data taking in 2022. First data collected in Run 3 allowed for an early test of these novel lines, showing an affordable rate and a good purity of collected candidates, suggesting their further exploitation in future physics data taking periods. 

The results reported here represent an important proof of concept in triggering at hadron-collider experiments, as the illustrated strategy can be  extended to other LLP decays, potentially impacting several SM and BSM studies, currently limited by available statistics. This approach also sets an interesting example beyond the LLP realm, of successful channel-specific trigger selections at the very first trigger level. This approach could be fruitfully applied to other kind of selections in future large luminosity scenarios at hadron colliders.
\bibliographystyle{JHEP}
\bibliography{main.bib}

\end{document}

%% file: lhcb-symbols-def.tex
\usepackage{xspace} 
\usepackage{upgreek}

\def\twotrks   {{\tt TwoTrackKs}\xspace}
\def\twoks   {{\tt TwoKs}\xspace}

\def\lhcb   {\mbox{LHCb}\xspace}
\def\atlas  {\mbox{ATLAS}\xspace}
\def\cms    {\mbox{CMS}\xspace}

\def\babar  {\mbox{BaBar}\xspace}
\def\belle  {\mbox{Belle}\xspace}

\def\velo   {VELO\xspace}

\def\MagUp {\mbox{\em Mag\kern -0.05em Up}\xspace}

\ifthenelse{\boolean{uprightparticles}}%
{

 \def\Peta        {\ensuremath{\upeta}\xspace}

 \def\Pmu         {\ensuremath{\upmu}\xspace}

 \def\Ppi         {\ensuremath{\uppi}\xspace}

 \def\Ppsi        {\ensuremath{\uppsi}\xspace}

 \def\PDelta      {\ensuremath{\Delta}\xspace}                 
 \def\PXi         {\ensuremath{\Xi}\xspace}                 
 \def\PLambda     {\ensuremath{\Lambda}\xspace}                 
 \def\PSigma      {\ensuremath{\Sigma}\xspace}                 
 \def\POmega      {\ensuremath{\Omega}\xspace}                 
 \def\PUpsilon    {\ensuremath{\Upsilon}\xspace}
 \let\oldPi\Pi
 \def\PPi         {\ensuremath{\oldPi}\xspace}

 \def\PB      {\ensuremath{\mathrm{B}}\xspace}                 
                  
 \def\PD      {\ensuremath{\mathrm{D}}\xspace}

 \def\PJ      {\ensuremath{\mathrm{J}}\xspace}                 
 \def\PK      {\ensuremath{\mathrm{K}}\xspace}

 \def\Pi      {\ensuremath{\mathrm{i}}\xspace}

 \def\Pp      {\ensuremath{\mathrm{p}}\xspace}

 \def\Ps      {\ensuremath{\mathrm{s}}\xspace}

 \def\thebaroffset{0.0em}
}
{

 \def\Peta        {\ensuremath{\eta}\xspace}

 \def\Pmu         {\ensuremath{\mu}\xspace}

 \def\Ppi         {\ensuremath{\pi}\xspace}

 \def\Ppsi        {\ensuremath{\psi}\xspace}                 
                  
 \mathchardef\PDelta="7101
 \mathchardef\PXi="7104
 \mathchardef\PLambda="7103
 \mathchardef\PSigma="7106
 \mathchardef\POmega="710A
 \mathchardef\PUpsilon="7107
 \mathchardef\PPi="7105
                  
 \def\PB      {\ensuremath{B}\xspace}                 
                  
 \def\PD      {\ensuremath{D}\xspace}

 \def\PJ      {\ensuremath{J}\xspace}                 
 \def\PK      {\ensuremath{K}\xspace}

 \def\Pi      {\ensuremath{i}\xspace}

 \def\Pp      {\ensuremath{p}\xspace}

 \def\Ps      {\ensuremath{s}\xspace}

 \def\thebaroffset{0.18em}
}
\newcommand{\offsetoverline}[2][\thebaroffset]{\kern #1\overline{\kern -#1 #2}}%

\makeatletter
\ifcase \@ptsize \relax
  \newcommand{\miniscule}{\@setfontsize\miniscule{4}{5}}
\or
  \newcommand{\miniscule}{\@setfontsize\miniscule{5}{6}}
\or
  \newcommand{\miniscule}{\@setfontsize\miniscule{5}{6}}
\fi
\makeatother

\DeclareRobustCommand{\optbar}[1]{\shortstack{{\miniscule (\rule[.5ex]{1.25em}{.18mm})}
  \\ [-.7ex] $#1$}}

\def\mup        {{\ensuremath{\Pmu^+}}\xspace}
\def\mun        {{\ensuremath{\Pmu^-}}\xspace} 

\def\squark    {{\ensuremath{\Ps}}\xspace}

\def\pion   {{\ensuremath{\Ppi}}\xspace}
\def\piz    {{\ensuremath{\pion^0}}\xspace}
\def\pip    {{\ensuremath{\pion^+}}\xspace}
\def\pim    {{\ensuremath{\pion^-}}\xspace}

\def\kaon    {{\ensuremath{\PK}}\xspace}

\def\KorKbar {\kern \thebaroffset\optbar{\kern -\thebaroffset \PK}{}\xspace}

\def\Kp      {{\ensuremath{\kaon^+}}\xspace}
\def\Km      {{\ensuremath{\kaon^-}}\xspace}

\def\KS      {{\ensuremath{\kaon^0_{\mathrm{S}}}}\xspace}

\def\D       {{\ensuremath{\PD}}\xspace}

\def\DorDbar {\kern \thebaroffset\optbar{\kern -\thebaroffset \PD}\xspace}
\def\Dz      {{\ensuremath{\D^0}}\xspace}

\def\Dp      {{\ensuremath{\D^+}}\xspace}
\def\Dm      {{\ensuremath{\D^-}}\xspace}

\def\DpDm    {\ensuremath{\Dp {\kern -0.16em \Dm}}\xspace}

\def\B       {{\ensuremath{\PB}}\xspace}

\def\BorBbar {\kern \thebaroffset\optbar{\kern -\thebaroffset \PB}\xspace}
\def\Bz      {{\ensuremath{\B^0}}\xspace}

\def\Bd      {{\ensuremath{\B^0}}\xspace}

\def\BdorBdbar {\kern \thebaroffset\optbar{\kern -\thebaroffset \Bd}\xspace}

\def\Bs      {{\ensuremath{\B^0_\squark}}\xspace}

\def\BsorBsbar {\kern \thebaroffset\optbar{\kern -\thebaroffset \Bs}\xspace}

\def\jpsi     {{\ensuremath{{\PJ\mskip -3mu/\mskip -2mu\Ppsi}}}\xspace}

\def\Y#1S{\ensuremath{\PUpsilon{(#1S)}}\xspace}

\def\proton      {{\ensuremath{\Pp}}\xspace}

\def\Lz          {{\ensuremath{\PLambda}}\xspace}

\def\LorLbar     {\kern \thebaroffset\optbar{\kern -\thebaroffset \PLambda}\xspace}

\def\to                 {\ensuremath{\rightarrow}\xspace}

\def\CP                {{\ensuremath{C\!P}}\xspace}

\newcommand{\ACP}{{\ensuremath{{\mathcal{A}}^{\CP}}}\xspace}

\def\AT#1     {\ensuremath{A_{\mathrm{T}}^{#1}}\xspace}           

\def\C#1      {\ensuremath{\mathcal{C}_{#1}}\xspace}                       
\def\Cp#1     {\ensuremath{\mathcal{C}_{#1}^{'}}\xspace}                    
\def\Ceff#1   {\ensuremath{\mathcal{C}_{#1}^{\mathrm{(eff)}}}\xspace}        
\def\Cpeff#1  {\ensuremath{\mathcal{C}_{#1}^{'\mathrm{(eff)}}}\xspace}       
\def\Ope#1    {\ensuremath{\mathcal{O}_{#1}}\xspace}                       
\def\Opep#1   {\ensuremath{\mathcal{O}_{#1}^{'}}\xspace}                    

\newcommand{\aunit}[1]{\ensuremath{\text{\,#1}}}       

\newcommand{\tev}{\aunit{Te\kern -0.1em V}\xspace}
\newcommand{\gev}{\aunit{Ge\kern -0.1em V}\xspace}
\newcommand{\mev}{\aunit{Me\kern -0.1em V}\xspace}
\newcommand{\kev}{\aunit{ke\kern -0.1em V}\xspace}
\newcommand{\ev}{\aunit{e\kern -0.1em V}\xspace}
 
\newcommand{\mevc}{\ensuremath{\aunit{Me\kern -0.1em V\!/}c}\xspace}
\newcommand{\gevc}{\ensuremath{\aunit{Ge\kern -0.1em V\!/}c}\xspace}
\newcommand{\mevcc}{\ensuremath{\aunit{Me\kern -0.1em V\!/}c^2}\xspace}
\newcommand{\gevcc}{\ensuremath{\aunit{Ge\kern -0.1em V\!/}c^2}\xspace}

\def\mm   {\aunit{mm}\xspace}

\def\nb {\aunit{nb}\xspace}
\def\invnb {\ensuremath{\nb^{-1}}\xspace}

\def\fb   {\ensuremath{\aunit{fb}}\xspace}
\def\invfb   {\ensuremath{\fb^{-1}}\xspace}

\def\sec  {\ensuremath{\aunit{s}}\xspace}

\def\mhz  {\ensuremath{\aunit{MHz}}\xspace}
\def\khz  {\ensuremath{\aunit{kHz}}\xspace}

\newcommand{\chisq}{\ensuremath{\chi^2}\xspace}
\newcommand{\chisqndf}{\ensuremath{\chi^2/\mathrm{ndf}}\xspace}
\newcommand{\chisqip}{\ensuremath{\chi^2_{\text{IP}}}\xspace}

\newcommand{\chisqvtxndf}{\ensuremath{\chi^2_{\text{vtx}}/\mathrm{ndf}}\xspace}

\def\gsim{{~\raise.15em\hbox{$>$}\kern-.85em
          \lower.35em\hbox{$\sim$}~}\xspace}
\def\lsim{{~\raise.15em\hbox{$<$}\kern-.85em
          \lower.35em\hbox{$\sim$}~}\xspace}

\def\pt         {\ensuremath{p_{\mathrm{T}}}\xspace}

\def\tell1  {TELL1\xspace}
\def\ukl1   {UKL1\xspace}

\newcommand{\lhcborcid}[1]{\href{https://orcid.org/#1}{\hspace*{0.1em}\raisebox{-0.45ex}{\includegraphics[width=1em]{figs/orcidIcon.pdf}}}}
